%% file: TOP-20-007_temp.tex
\pdfoutput=1
\documentclass[11pt,twoside,a4paper,cmspaper,final,collab]{cms-tdr}
\def\svnVersion{9195c8a}\def\svnDate{2022/07/13}\def\cmsCernNoTag{CERN-EP-2021-201}\def\cmsCernDate{\today}\def\cmsMessage{Published in Physical Review Letters as \href{http://dx.doi.org/10.1103/PhysRevLett.129.032001}{\doi{10.1103/PhysRevLett.129.032001}.}}
\begin{document}\cmsNoteHeader{TOP-20-007}

\ifthenelse{\boolean{cms@external}}{\providecommand{\cmsLeft}{upper\xspace}}{\providecommand{\cmsLeft}{left\xspace}}
\ifthenelse{\boolean{cms@external}}{\providecommand{\cmsRight}{lower\xspace}}{\providecommand{\cmsRight}{right\xspace}}

\newcommand{\Hgg}{\ensuremath{\PH \to \PGg\PGg}\xspace}
\newcommand{\tqH}{\ensuremath{\PQt \to \PH\PQq}\xspace}
\newcommand{\tuH}{\ensuremath{\PQt \to \PH\PQu}\xspace}
\newcommand{\tcH}{\ensuremath{\PQt \to \PH\PQc}\xspace}
\newcommand{\brHqt}{\ensuremath{\mathcal{B}(\PQt \to \PH\PQq)}\xspace}
\newcommand{\brHut}{\ensuremath{\mathcal{B}(\PQt \to \PH\PQu)}\xspace}
\newcommand{\brHct}{\ensuremath{\mathcal{B}(\PQt \to \PH\PQc)}\xspace}
\newcommand{\mgg}{\ensuremath{m_{\PGg\PGg}}\xspace}
\newcommand{\ttbarr}{\ensuremath{\ttbar + \text{jets}}\xspace}
\newcommand{\ttgammagamma}{\ensuremath{\ttbar + \PGg\PGg}\xspace}
\newcommand{\ttgamma}{\ensuremath{\ttbar + \PGg}\xspace}
\newcommand{\gjet}{\ensuremath{\PGg + \text{jets}}\xspace}
\newcommand{\dipho}{\ensuremath{\PGg\PGg + \text{jets}}\xspace}
\newcommand{\vgamma}{\ensuremath{\PV + \PGg}\xspace}
\newcommand{\ww}{\ensuremath{\PW\PW}\xspace}
\newcommand{\zz}{\ensuremath{\PZ\PZ}\xspace}
\newcommand{\wz}{\ensuremath{\PW\PZ}\xspace}
\newcommand{\ttW}{\ensuremath{\ttbar\PW}\xspace}
\newcommand{\ttZ}{\ensuremath{\ttbar\PZ}\xspace}
\newcommand{\tg}{\ensuremath{\PQt + \PGg + \text{jets}}\xspace}
\newcommand{\ttH}{\ensuremath{\ttbar\PH}\xspace}
\newcommand{\WH}{\ensuremath{\PW\PH}\xspace}
\newcommand{\ZH}{\ensuremath{\PZ\PH}\xspace}
\newcommand{\VBF}{\ensuremath{\mathrm{VBF}}\xspace}
\newcommand{\bbH}{\ensuremath{\bbbar\PH}\xspace}
\newcommand{\tHq}{\ensuremath{\PQt\PH\PQq}\xspace}
\newcommand{\tHW}{\ensuremath{\PQt\PH\PW}\xspace}
\newcommand{\ggH}{\ensuremath{\cPg\cPg\PH}\xspace}
\newcommand{\mH}{\ensuremath{m_{\PH}}\xspace}
\newcommand{\mt}{\ensuremath{m_{\PQt}}\xspace}
\newcommand{\mggj}{\ensuremath{m_{\PGg\PGg\mathrm{j}}}\xspace}
\newcommand{\mjjj}{\ensuremath{m_{\mathrm{jjj}}}\xspace}

\cmsNoteHeader{TOP-20-007}
\title{Search for flavor-changing neutral current interactions of the top quark and Higgs boson in final states with two photons in proton-proton collisions at \texorpdfstring{$\sqrt{s}=13\TeV$}{sqrt(s) = 13 TeV}}

\date{\today}

\abstract{Proton-proton interactions resulting in final states with two photons are studied in a search for the signature of flavor-changing neutral current interactions of top quarks~(\PQt) and Higgs bosons~(\PH). The analysis is based on data collected at a center-of-mass energy of 13\TeV with the CMS detector at the LHC, corresponding to an integrated luminosity of 137\fbinv. No significant excess above the background prediction is observed. Upper limits on the branching fractions ($\mathcal{B}$) of the top quark decaying to a Higgs boson and an up (\PQu) or charm quark (\PQc) are derived through a binned fit to the diphoton invariant mass spectrum. The observed (expected) 95\% confidence level upper limits are found to be $0.019\,(0.031)\%$ for $\mathcal{B}(\PQt \to \PH\PQu)$ and $0.073\,(0.051)\%$ for $\mathcal{B}(\PQt \to \PH\PQc)$. These are the strictest upper limits yet determined.}

\hypersetup{
pdfauthor={CMS Collaboration}, 
pdftitle={Search for flavor-changing neutral current interactions of the top quark and Higgs boson in final states with two photons in proton-proton collisions at sqrt(s)=13 TeV},
pdfsubject={CMS},
pdfkeywords={CMS,  flavor-changing neutral currents, top quarks, Higgs to diphoton}} 

\maketitle 
Flavor-changing quark decays mediated by neutral currents (FCNCs) 
are forbidden at tree level in the standard model (SM).
They may proceed at higher orders in the perturbative expansion;
however, these rates are heavily suppressed by the Glashow--Iliopoulos--Maiani mechanism~\cite{Glashow:1970gm} or Cabibbo--Kobayashi--Maskawa unitarity constraints~\cite{Kobayashi:1973fv}.
The SM branching fractions for
the decay of a top quark (\PQt) into a Higgs boson (\PH) and up quark (\PQu),
\tuH, or charm quark (\PQc), \tcH, are expected to be
$\mathcal{O}(10^{-17})$ and $\mathcal{O}(10^{-15})$, respectively~\cite{Eilam:1990zc,Mele_1998,AguilarSaavedra:2004wm,Zhang_2013}, well below the current sensitivity of the LHC experiments~\cite{Aaboud:2018oqm}.
Thus, any observation of a \tqH FCNC interaction would be an unambiguous sign of new physics.
Here, the symbol \PQq denotes either an up or charm quark.

In many scenarios of physics beyond the SM, the \tqH branching fractions are enhanced by many orders of magnitude beyond the SM values.
Notable beyond-the-SM models leading to enhanced FCNC interactions include those of warped extra dimensions~\cite{Azatov_2009}, 
composite Higgs boson models~\cite{Azatov_2014}, 
two-Higgs doublet models (2HDM)~\cite{Bejar:2000ub,Guasch_1999,Cao_2007,Cao_2014}, 
including supersymmetric models with R-parity violation~\cite{Eilam_2001},
and quark-singlet models~\cite{AguilarSaavedra:2002kr}.
While these scenarios lead to sizable FCNC interactions for a variety of neutral mediators other than the Higgs boson, including the \PZ boson, photon, and gluon, some of the most significant enhancements are found for \tqH interactions.
The \tcH interaction in particular can be enhanced in 2HDM models~\cite{Hou:1991un}, including scenarios of flavor-violating Yukawa couplings~\cite{Chen:2013qta} or 2HDM for the top quark~\cite{Baum:2008qm,Kao:2011aa,Altunkaynak:2015twa}.

\begin{figure*} [htb!]
    \centering
    \includegraphics[width=0.7\textwidth]{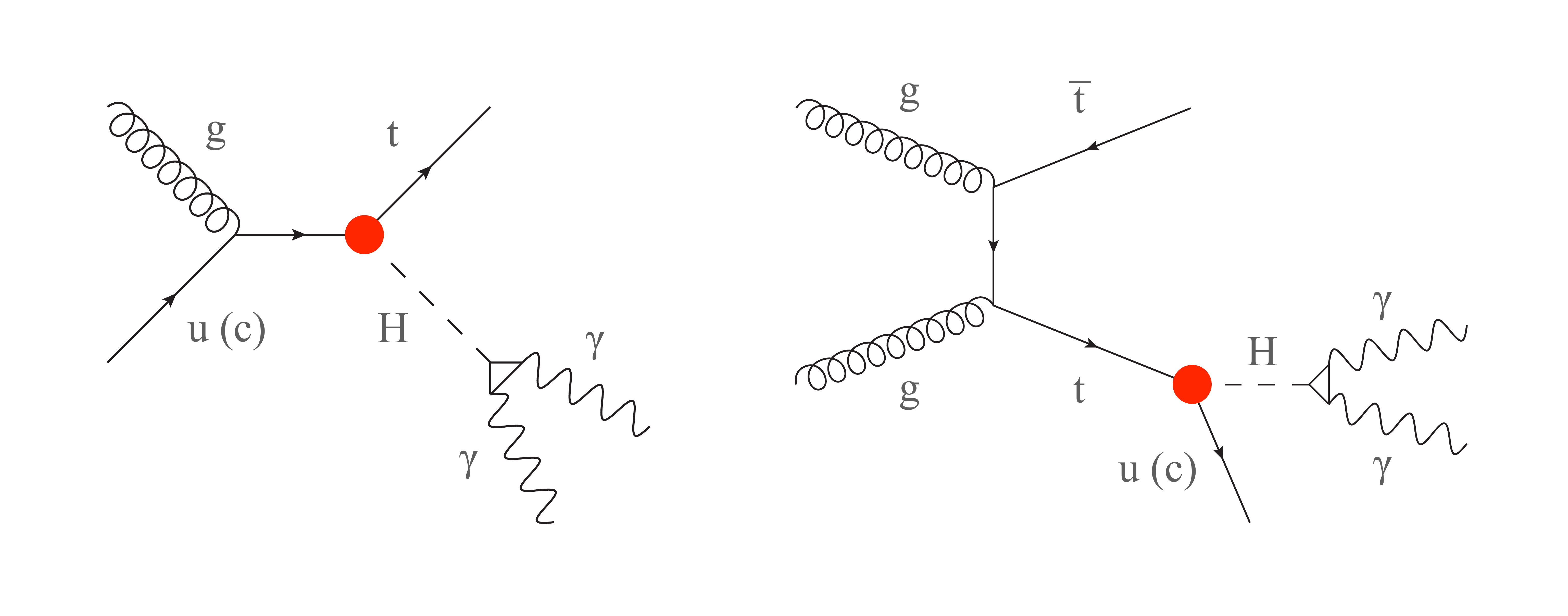}
    \caption{Representative Feynman diagrams for the production modes considered: FCNC associated production of a single top quark with a Higgs boson (ST, left), and \ttbar production with the FCNC decay of the top quark to a Higgs boson and an up or charm quark (TT, right). The Higgs boson decay to two photons is considered. The FCNC vertex in each process is denoted with a red circle.}	
    \label{fig:feyn}
\end{figure*}

Recent searches for FCNC interactions of the top quark and Higgs boson were performed by the ATLAS~\cite{Aaboud:2017mfd,Aaboud:2018pob,Aaboud:2018oqm} and CMS~\cite{Sirunyan:2017uae} Collaborations, with Ref.~\cite{Aaboud:2018oqm} providing the strictest experimental upper limits on the \tuH and \tcH branching fractions at $0.12$ and $0.11\%$, respectively.
This Letter reports improved upper limits on these two branching fractions, exploiting both the associated production of a single top quark with a Higgs boson via an up or charm quark (ST production mode) and the decay of a top quark to a Higgs boson and an up or charm quark in \ttbar production (TT production mode), as shown in Fig.~\ref{fig:feyn}, where the \Hgg decay is considered. 

The results are based on the analysis of proton-proton ($\Pp\Pp$) collisions at a center-of-mass energy of
$13$\TeV, targeting the \Hgg decay mode.  The data were
collected with the CMS detector at the LHC between 2016--2018, and correspond to an integrated luminosity of 137\fbinv.
Tabulated results are provided in the HEPData record for this analysis~\cite{hepdata}.

The central feature of the CMS apparatus is a superconducting solenoid of 6\unit{m} internal diameter, providing a magnetic field of 3.8\unit{T}.
Within the solenoid volume are a silicon pixel and strip tracker, a lead tungstate crystal electromagnetic calorimeter (ECAL), and a brass and scintillator hadron calorimeter, each composed of a barrel and two endcap sections.
Forward calorimeters extend the pseudorapidity ($\eta$) coverage provided by the barrel and endcap detectors. Muons are measured in gaseous detectors embedded in the steel flux-return yoke outside the solenoid.
A more detailed description of the CMS detector, together with a definition of the coordinate system used and the relevant kinematic variables, can be found in Ref.~\cite{Chatrchyan:2008zzk}.

The signal and background processes are simulated using several Monte Carlo (MC) programs.
The signal samples corresponding to ST and TT production modes are simulated at leading order (LO)
with \MGvATNLO 2.4.2 (ST) and 2.6.0 (TT)~\cite{Alwall:2014hca}.
Both the signal and background samples are interfaced with \PYTHIA 8.205~\cite{Sjostrand:2014zea} for parton showering, fragmentation, and hadronization.
The underlying event is also modeled with \PYTHIA, with the CUETP8M1~\cite{Khachatryan:2015pea} and CP5~\cite{Sirunyan:2019dfx} tunes used for the simulation of 2016 and 2017/2018 data, respectively.
The parton distribution functions (PDFs) are taken from the NNPDF~3.0~\cite{NNPDF3} set for simulation of 2016 data and the NNPDF~3.1~\cite{NNPDF31} set for simulation of 2017/2018 data.

The FCNC interactions, including those with the Higgs boson as a mediator,
can be described within the effective field theory framework in terms of dimension-six operators added to the SM Lagrangian~\cite{Buchmuller:1985jz,Grzadkowski:2010es}.
The coefficients of these operators are best constrained by combining the results of the FCNC processes in which the Higgs boson, Z boson, photon, and gluon are the FCNC mediators.
To perform this combination, it is important to have the most sensitive analysis for each individual process.
In this paper, the theoretical interpretation of the FCNC interactions with the Higgs boson as the mediator is done by using the following effective Lagrangian,
which is equivalent to the effective field theory approach at LO,
\begin{linenomath}
    \begin{equation}
            \mathcal{L} = \sum_{\PQq=\PQu, \PQc} \frac{g}{\sqrt{2}} \PAQt \kappa_{\PH\PQq\PQt} \Big( F_{\PH\PQq}^{\mathrm{L}} P_{\mathrm{L}} + F_{\PH\PQq}^{\mathrm{R}} P_{\mathrm{R}} \Big) \PQq\PH + \text{h.c.}, \label{eqn:lagrangian}
    \end{equation}
\end{linenomath}
where $g$ is the weak coupling constant, $P_{\mathrm{L}}$ and $P_{\mathrm{R}}$ are chirality projectors in spin space, $\kappa_{\PH\PQq\PQt}$ is the effective coupling constant, with $\PQq = \PQu$ or $\PQc$, and $F_{\PH\PQq}^{\mathrm{L}}$ and $F_{\PH\PQq}^{\mathrm{R}}$ are left- and right-handed complex chiral parameters satisfying a unitarity constraint $\abs{F_{\PH\PQq}^{\mathrm{L}}}^2 + \abs{F_{\PH\PQq}^{\mathrm{R}}}^2 = 1$. 
The effective Lagrangian is implemented in the \textsc{FeynRules} package~\cite{Alloul:2013bka}, with the universal \textsc{FeynRules} output~\cite{Degrande:2011ua} used to generate the model.
The complex chiral parameters are set to $F_{\PH\PQq}^{\mathrm{L}} = 1$ and $F_{\PH\PQq}^{\mathrm{R}} = 0$.
Up to two additional partons are generated at matrix element level for the TT production mode.
No additional partons are considered for the ST production mode to avoid overlap between the ST and TT modes.
The hard-process simulation is interfaced with parton shower modeling using the MLM~\cite{Alwall:2007fs} matching prescription.
Signal samples are generated in two scenarios, assuming exactly one nonzero coupling of unity: $\kappa_{\PH\PQu\PQt}$ or $\kappa_{\PH\PQc\PQt} = 1$.
The cross section of the ST production mode is calculated with \MADGRAPH at LO precision and equals 72.6 (10.0)\unit{pb} for the scenarios of $\kappa_{\PH\PQu\PQt} (\kappa_{\PH\PQc\PQt}) = 1$.
The \ttbar cross section is taken as 832\unit{pb}, from calculation with the \TOPpp program~\cite{Czakon:2011xx} at next-to-next-to-LO (NNLO) in perturbative quantum chromodynamics (QCD), which includes soft-gluon resummation to next-to-next-to-leading-logarithmic order.
The cross section times branching fraction for the TT FCNC production mode depends on the branching fraction of \tuH and \tcH, which is 0.144 (as calculated by \MADGRAPH at LO) when assuming a coupling of $\kappa_{\PH\PQu\PQt} = \kappa_{\PH\PQc\PQt} = 1$.
The effective coupling constant and branching fraction are related by 
\begin{linenomath}
    \begin{equation}
			\kappa_{\PH\PQq\PQt}^2 = \mathcal{B}(\PQt \to \PH \PQq) \frac{\Gamma_\PQt}{\Gamma_{\PH\PQq\PQt}}, \label{eqn:coupling_vs_bf}
    \end{equation}
\end{linenomath}
where $\Gamma_\PQt$ is the full width of the top quark~\cite{Gao:2012ja}, and $\Gamma_{\PH\PQq\PQt}$ is the partial width of the anomalous $\PQt \to \PH \PQq$ decay. 

Calculations of the top quark transverse momentum (\pt) distribution at LO and next-to-LO (NLO) are known to show disagreement with the observed spectrum~\cite{Kidonakis:2012rm,Czakon:2015owf,Czakon:2017wor,Catani:2019hip}.
Since the effect may be partially due to missing higher-order calculations, the simulated top \pt distribution in TT signal events is corrected at generator level to match the NNLO QCD + NLO electroweak prediction~\cite{Czakon:2017wor}.

Background processes with a Higgs boson decaying to two photons (resonant backgrounds), including \ggH, vector boson fusion (\VBF), \WH, \ZH, \tHq, \tHW, \ttH, and \bbH production, are modeled with \MGvATNLO 2.4.2 at NLO~\cite{Alwall:2014hca} in QCD, with cross sections and decay branching fractions
taken from Ref.~\cite{deFlorian:2016spz}.
Additional samples for SM Higgs boson processes are generated with \POWHEG~2.0~\cite{Nason:2004rx,Frixione:2007vw,Alioli:2010xd,Hartanto:2015uka}
at NLO in QCD.  They are used to increase the size of the SM data sample needed for training the multivariate discriminants described below.

The \MGvATNLO program is also used to simulate most background processes without a Higgs boson (nonresonant backgrounds).
These include \ttgammagamma, \ttgamma, \ttbarr, \gjet, \vgamma, Drell--Yan, diboson, and $\PQt + \PV$ production, where \PV is a \PW or \PZ boson.
The diphoton background (\dipho) is simulated with \SHERPA
2.2.4~\cite{10.21468/SciPostPhys.7.3.034}, which
includes tree-level processes with up to three additional jets, as well as box processes at LO accuracy.
Simulation of nonresonant background samples is used for developing and optimizing the analysis, but this background is estimated from data
in the fits used to extract a possible signal contribution.
Finally, for both signal and background samples, the detector response is modeled with the \GEANTfour package \cite{Agostinelli:2002hh}.

The CMS trigger~\cite{Khachatryan:2016bia} selects events with two photon candidates satisfying a loose calorimetric
identification~\cite{Sirunyan:2018ouh}, and asymmetric photon \pt thresholds
of 30 and 18 (22)\GeV for the data collected during 2016 (2017/2018).
The trigger efficiency is $>$95\% and is measured with the ``tag-and-probe'' method~\cite{CMS:2011aa}.

The particle-flow (PF) algorithm~\cite{CMS-PRF-14-001}
reconstructs individual particles (photons,
charged and neutral hadrons, muons, and electrons)
by combining information from all subdetectors.
Jets are built from PF particles using the anti-\kt algorithm~\cite{Cacciari:2008gp,Cacciari:2011ma}
with a distance parameter of 0.4.
The missing transverse momentum ($\ptvecmiss$)~\cite{Sirunyan:2019kia} is defined as the negative vector \pt sum of all PF
particles. Its magnitude is referred to as $\ptmiss$.
The primary $\Pp\Pp$ interaction vertex is taken as the vertex with the largest value of summed physics-object $\pt^{2}$.
Charged hadrons originating from other $\Pp\Pp$ interactions in the same or nearby bunch crossings are removed
from the analysis. Jets from the hadronization of bottom
quarks (\PQb jets) are tagged by an algorithm
based on the score from a deep neural network~\cite{Sirunyan:2017ezt} called \textsc{DeepJet}~\cite{Bols:2020bkb,CMS-DP-2018-058}.

Higgs boson candidates are built from pairs of photon candidates.
Photon candidates are reconstructed from energy clusters in the ECAL not linked to
charged particle tracks (with the exception of converted photons).
The photon energy is corrected for the containment of electromagnetic showers
in the clustered crystals and the energy losses of converted photons
with a multivariate regression technique based on simulation~\cite{Sirunyan:2018ouh}.
The ECAL energy scale in data is corrected based on studies of
$\PZ\to\EE$ events.
The offline diphoton selection criteria are
similar to, but more stringent than, those used in the trigger~\cite{Sirunyan:2018ouh}.

Photons are further required to satisfy a loose identification (photon ID~\cite{Sirunyan:2018ouh})
criterion based on a boosted decision tree (BDT) classifier trained to separate photons from jets.
In simulation, inputs to the photon ID such as shower shape and isolation variables are corrected with
a chained quantile regression method~\cite{DBLP:journals/corr/abs-1211-6581} based on studies of $\PZ\to\EE$ events.
The cumulative distribution function of each variable in simulation is corrected with a set of BDTs
that take the previously corrected features as inputs.
This method ensures that both the individual distributions and correlations between variables in the MC simulation
are corrected to match those in data.

After the selection described above, the diphoton invariant mass (\mgg) is required to be in the range 100--180\GeV.
The upper bound of 180\GeV is selected to ensure a sufficient number of events for the procedures used to model the nonresonant background and to validate the modeling of input features to the BDTs, both described later. 
We additionally impose mass-dependent photon \pt requirements of $\pt/\mgg> 1/3$ and ${>}1/4$ for the highest \pt (leading) and second-highest \pt (subleading) photons, respectively.
Events are next divided into two mutually exclusive channels.
The leptonic channel preselection is aimed at selecting events with a leptonically decaying top quark,
and requires the presence of ${\geq}1$ jet with $\pt > 25$\GeV and $\abs{\eta} < 2.4$,
${\geq}1$ isolated lepton (\Pe or \PGm) with $\abs{\eta} < 2.4$ and $\pt > 10\,(20)$\GeV for electrons (muons)~\cite{Sirunyan:2018fpa}.
The hadronic channel preselection targets events with a hadronically decaying top quark,
requiring at least three jets, 
of which at least one is identified as a \PQb jet, and no isolated leptons (\Pe/\PGm).

Dedicated BDT discriminants are employed in each channel to distinguish signal (ST and TT production modes) from background events.
Separate BDTs are trained for each of the two nonzero $\kappa_{\PH\PQq\PQt}$ coupling scenarios, each of the two channels,
and each of the two primary categories of SM background, resonant (BDT-res) and nonresonant (BDT-nonres).
This results in eight BDTs in total: one for each coupling ($\kappa_{\PH\PQu\PQt}$ or $\kappa_{\PH\PQc\PQt}$), channel (hadronic or leptonic), and background (resonant or nonresonant).
The BDTs are trained with the \textsc{xgboost} \cite{xgboost} framework
on MC samples of signal and background processes, with one exception as noted below.
The nonresonant background MC samples include \gjet, \dipho, \ttbarr, \ttgamma, \ttgammagamma, and \vgamma processes,
as well as a variety of other rarer backgrounds (designated in figures as ``Other''), including \ttZ, \ttW, \ww, \wz, \zz, and \tg.
The resonant background MC samples include \ttH, \WH, \ZH, \VBF, \ggH, \tHq, and \tHW.
In the training, all backgrounds are normalized to their SM cross sections.

The dominant backgrounds after the hadronic channel preselection are the multijet and \gjet processes, where at least one jet is misidentified as a photon.
To improve the description of these backgrounds for training the machine learning algorithms,
they are modeled using a sample of
data events 
in which one photon candidate fails the photon ID requirement,
following the procedure of Ref.~\cite{Sirunyan:2020sum}.
For this procedure, the photon ID value of the photon candidate failing the photon ID requirement
is replaced by a value drawn from the MC photon ID distribution for misidentified jets passing the photon ID requirement.
This sample of events from data, denoted with (\PGg) + jets, is used instead of the simulated \gjet sample in the hadronic nonresonant background BDT training.

Input features of both BDT-res and BDT-nonres include the kinematic properties of the physics objects:
jets, leptons, photons and diphotons (excluding \mgg),
jet and lepton multiplicities, \ptmiss,
\PQb tagging scores of jets from the \textsc{DeepJet} algorithm,
and the output of the photon ID BDT for both photons.
The \pt of the individual photons and diphotons are normalized by \mgg to ensure the inputs to the BDTs cannot be used to calculate \mgg.
The output of a variety of algorithms aimed at reconstructing top quarks in each event are also used as input features to the BDTs, as detailed below.

First, a set of mass variables is used for events with at least four jets in the hadronic channel.
The invariant mass of the diphoton candidate plus the light jet from \tqH decays (\mggj) should be consistent with the top quark mass (\mt), as should the invariant mass of the three jets (\mjjj) from the hadronically decaying top quark candidate.
The mass variables \mggj and \mjjj are constructed from three of the leading four jets by choosing the combination that minimizes the quantity $\Delta M = \abs{\mggj - \mt} + \abs{\mjjj - \mt}$.
Second, kinematic reconstructions of top quarks and their decay products for both hadronic and leptonic decays are performed,
with the reconstruction for hadronic decays similar to that in Ref.~\cite{Sirunyan:2017uae} and the reconstruction for leptonic decays similar to that in Ref.~\cite{CMS:2018hfk}.
The kinematic properties of the reconstructed top quarks and their decay products are used as input features to the BDTs.
Third, a set of neural networks is trained with the TMVA package~\cite{tmva} on MC samples of signal processes to identify the jets and leptons that originate from the top quark(s) decays.
Permutations of jets and leptons matched to a top quark decay are considered as signal, while all incorrect permutations are considered as background.
Separate neural networks are trained for each signal production mode (ST and TT) and each channel (hadronic and leptonic).

\begin{figure*} [htb!]
	\centering
    \includegraphics[width=0.445\textwidth]{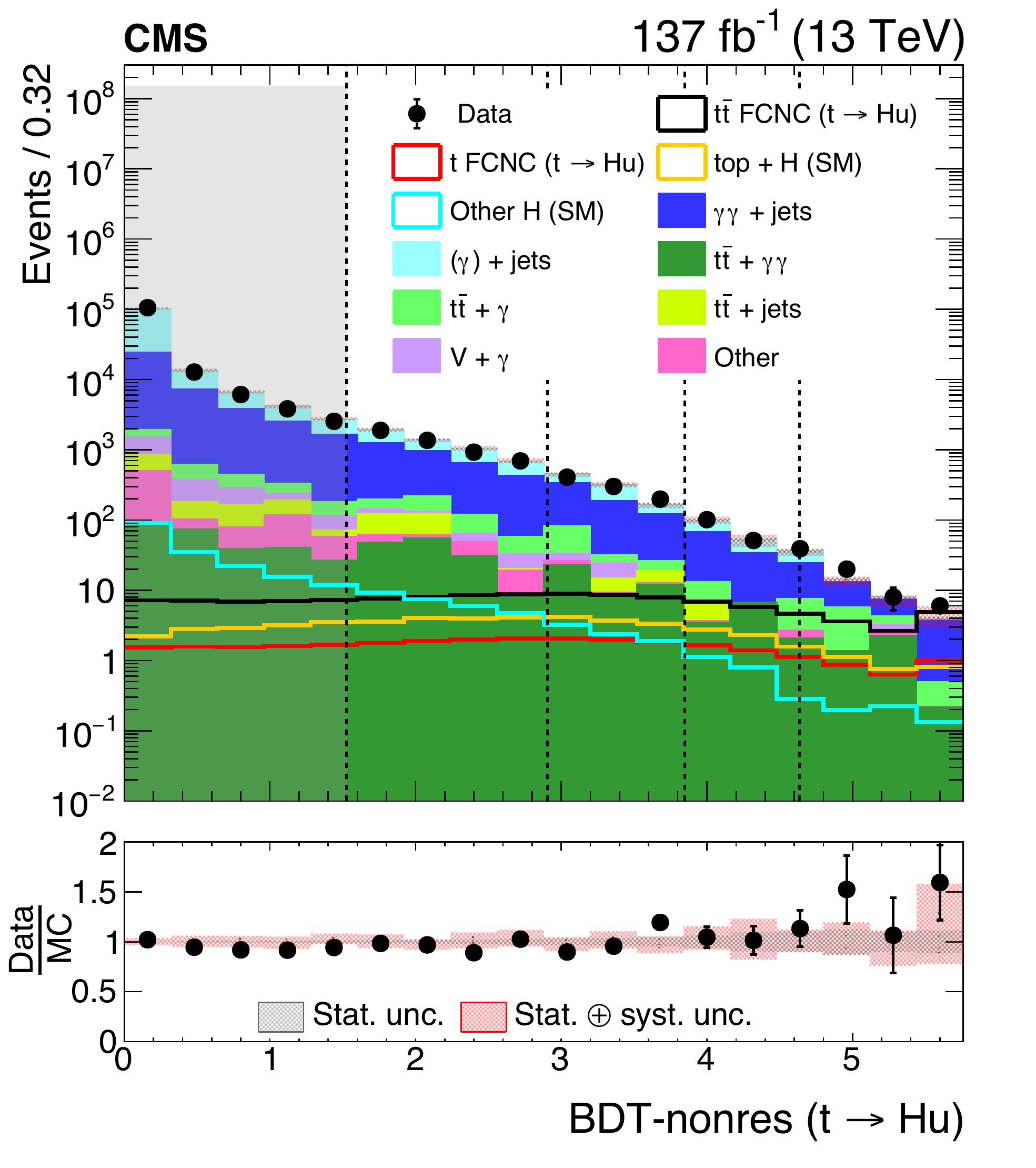} 
    \includegraphics[width=0.515\textwidth]{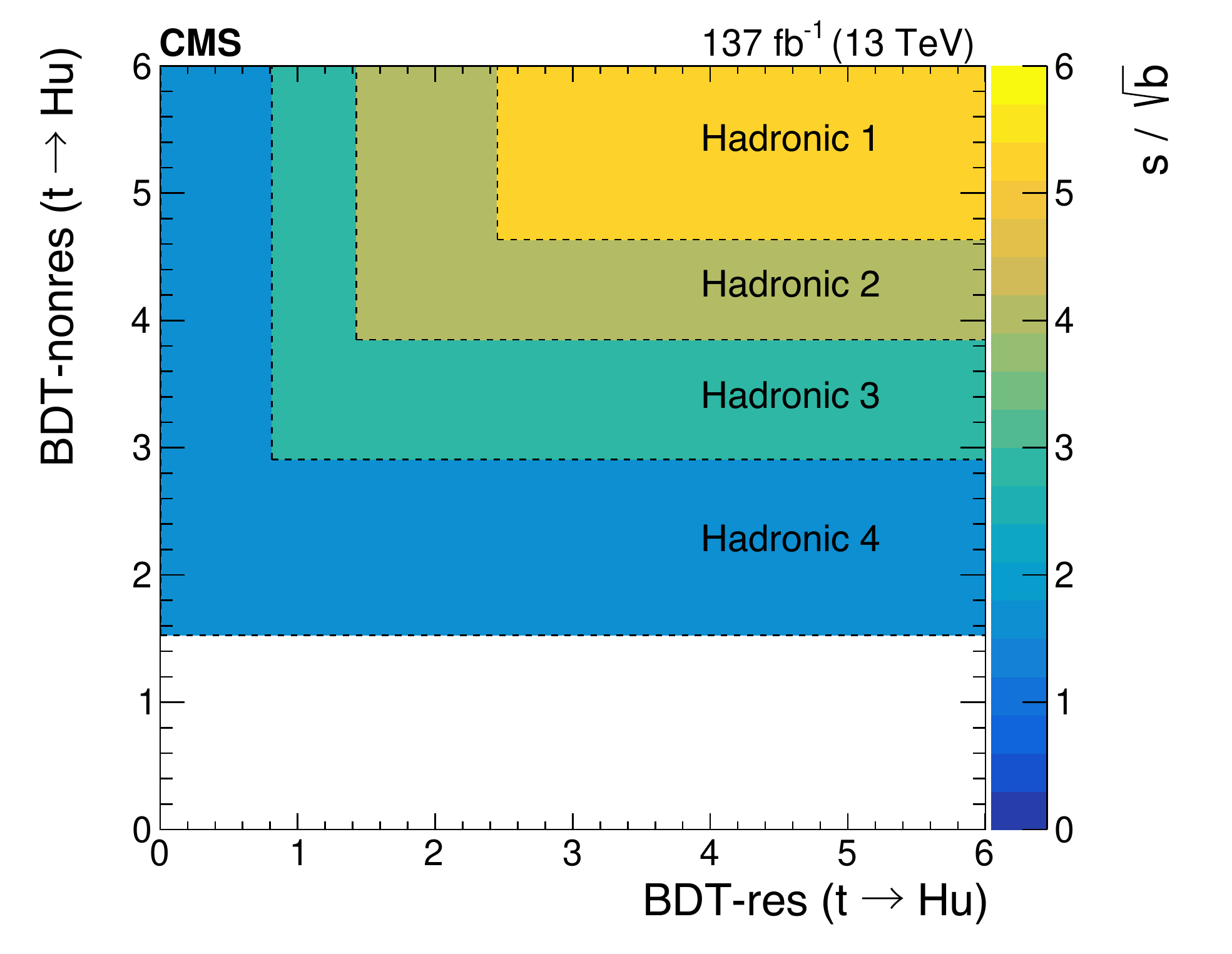} 
    \caption{Left: distribution of the BDT-nonres scores used for the hadronic event categorization targeting \tuH FCNC interactions from data (points) and predictions from simulation (colored histograms). 
             The ``Other'' category includes contributions from \ttZ, \ttW, \ww, \wz, \zz, and \tg.
             The ``top + H'' category includes \ttH, \tHq, and \tHW, while the ``Other H'' category includes \ggH, \VBF, \WH, and \ZH.
             The nonresonant background histograms are stacked, while the two signal, ``Other H'', and ``top + H'' distributions are shown separately.
             Boundaries defining event categories are indicated with dotted lines in the upper panel.
             Events in the grey shaded region are not considered in the analysis.
             The lower panel shows the ratio of the data to the sum of the nonresonant background predictions.
             The simulated signal and background distributions are normalized to the integrated luminosity of the data, assuming a coupling of unity for the signal.
             Statistical and total (statistical $\oplus$ systematic) background uncertainties are represented by the grey- and red-shaded bands, respectively.
             The (\PGg) + jets sample of multijet and \gjet events from data is not assigned a systematic uncertainty, as described in the text.
             Right: hadronic event categorization targeting \tuH FCNC interactions from requirements placed on BDT-nonres (horizontal dotted lines) and BDT-res (vertical dotted lines). The color denotes the $s/\sqrt{b}$ of each category. 
    }
    \label{fig:bdt_had_hut}
\end{figure*}

The modeling of the input features is validated by comparing their distributions in data and simulation for events passing the preselection and having \mgg in the sidebands, defined as the $\mgg$ ranges of 100--120 or 130--180\GeV. 
The resulting BDT scores are also validated in the same way, as shown in Fig.~\ref{fig:bdt_had_hut} for the \tuH search.
We note that the sample of events from data used to model the multijet and \gjet processes is only used in the BDT training and optimization and does not enter the fits used to extract possible FCNC signals.
Consequently, no systematic uncertainty is considered for the sample, despite that is not a perfect representation of these processes.

For each scenario of a nonzero FCNC coupling ($\kappa_{\PH\PQu\PQt}$ or $\kappa_{\PH\PQc\PQt} = 1$), and for each channel (hadronic or leptonic), events are either removed from consideration or assigned to categories.
The categories are defined by ranges of the BDT-nonres and BDT-res scores, shown, for example, with the four categories in the \tuH search in the hadronic channel indicated by the horizontal and vertical dotted lines in Fig.~\ref{fig:bdt_had_hut} (right).
In the same fashion, three categories are defined for the \tuH search in the leptonic channel, using the leptonic BDT-nonres and BDT-res scores.
The procedure is repeated for the \tcH search, resulting in a total of seven categories for each of the \tuH and \tcH searches.
Each resulting set of seven \mgg distributions is then fitted simultaneously to extract a possible FCNC signal. 
The categories for the \tuH (\tcH) search contain between 13--26 (2--5)\% of the total signal from ST production.

The expected \mgg distributions of signal and resonant background events are modeled using the sum of a double-sided Crystal Ball function~\cite{CrystalBallRef} and a
Gaussian function.
The models are derived from simulation for signal as well as each type of resonant background (\ggH, \VBF, \WH, \ZH, \tHq, \tHW, \ttH, and \bbH),
with the Higgs boson mass (\mH) fixed to its most precisely measured value of 125.38\unit{GeV}~\cite{CMS:2019drq}.
The nonresonant background is modeled directly from data, using the discrete profiling method~\cite{Dauncey:2014xga},
in which the systematic uncertainty associated with the choice of analytic function used to model the \mgg distribution is treated as a discrete nuisance parameter.
All sources of experimental and theoretical systematic uncertainties are treated as nuisance parameters.

The total \ttbar cross section uncertainty is taken as 6\%, estimated from the uncertainty in the \ttbar NNLO cross section, due to variation of the factorization ($\mu_F$) and renormalization ($\mu_R$) scales, the PDFs, and strong coupling
constant \alpS~\cite{Butterworth:2015oua,Martin:2009bu,Gao:2013xoa,Ball:2012cx}.
We assign an uncertainty in the ST signal production mode cross section of 30\% that is typically attributed to the missing higher-order corrections in the LO generation used in the analysis.
The typical effect of varying $\mu_F$ and $\mu_R$ on the shapes of the BDT-nonres and BDT-res distributions is around 1\,(10)\% for the TT\,(ST) signal production mode.
The uncertainties in the cross sections of the resonant background processes are estimated by varying $\mu_F$ and $\mu_R$, PDFs, and $\alpS$~\cite{deFlorian:2016spz}.

The dominant experimental uncertainties are those related to the \PQb jet and photon identifications,
the integrated luminosity~\cite{CMS-LUM-17-003,CMS-PAS-LUM-17-004,CMS-PAS-LUM-18-002},
the jet energy scale and resolution,
reconstruction of \ptvecmiss, 
and the preselection and trigger efficiencies.
The impact of each of these uncertainties on the final upper limits is $\leq$5\%. 

No significant excess above the background prediction is observed in any of the categories.
Binned fits of the \mgg distributions are performed simultaneously in each set of seven categories (14 total) to extract the 95\% confidence level (\CL) upper limits on \brHut and \brHct.
The derivation of upper limits assumes one nonzero coupling at a time and uses the modified frequentist approach for confidence levels (\CLs technique), with the LHC profile likelihood ratio as a test statistic~\cite{Junk:1999kv, Read_2002} in the asymptotic approximation~\cite{Cowan:2010js}.

\begin{figure} [htb!]
    \centering
    \includegraphics[width=0.48\textwidth]{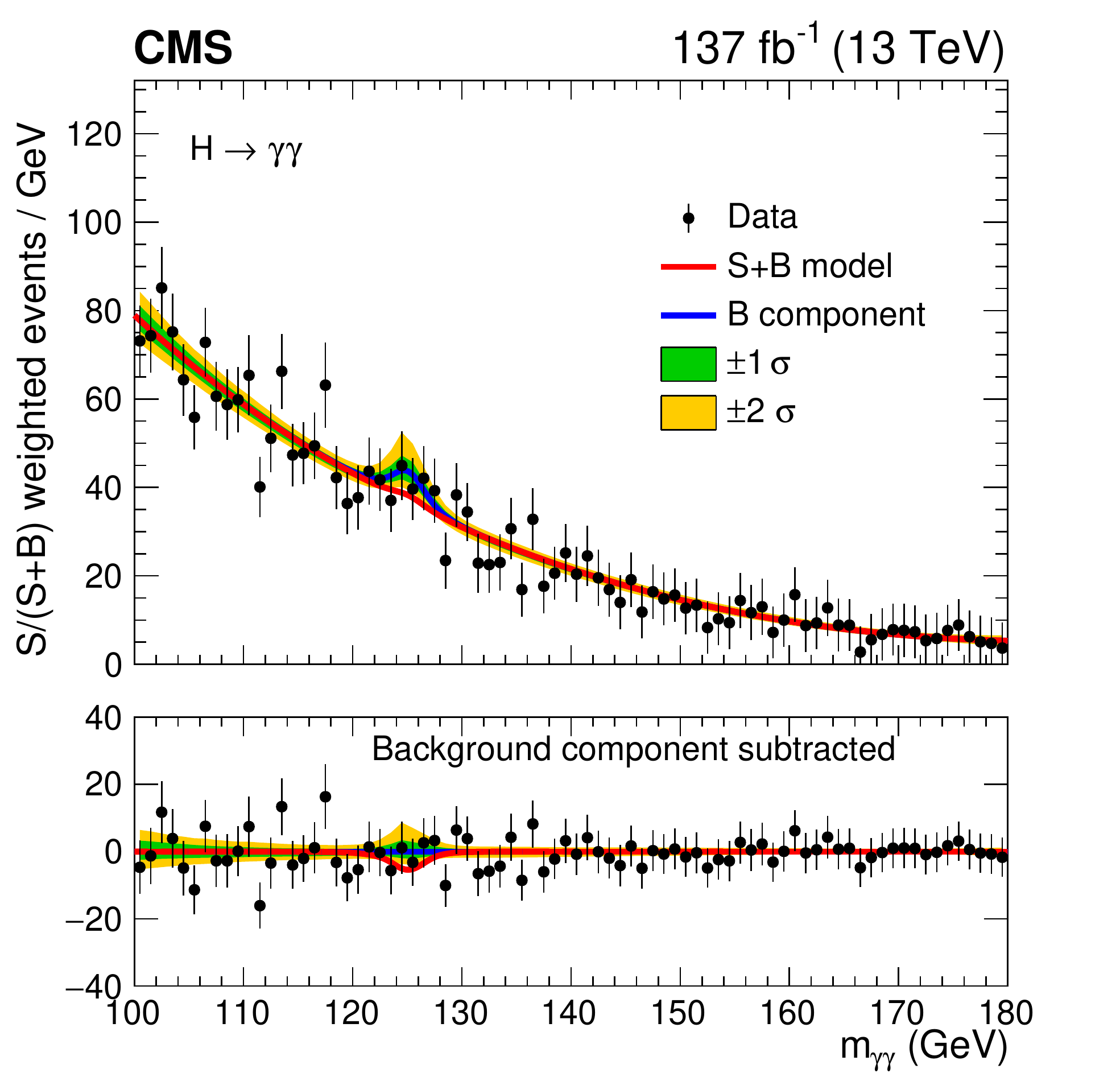}
    \includegraphics[width=0.48\textwidth]{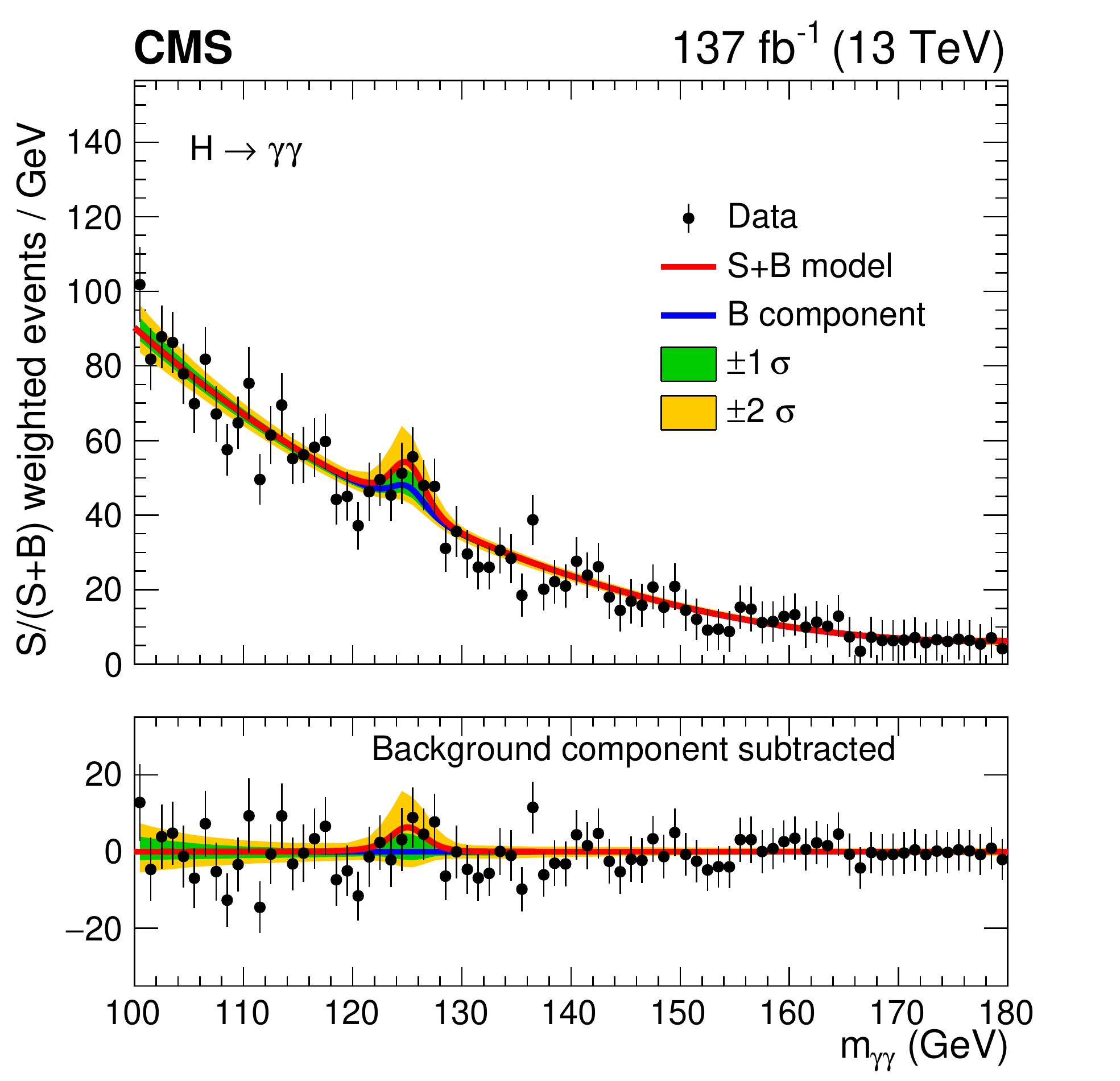}
    \caption{The diphoton invariant mass distribution for the selected events from data (black points),
             and the results of the fits to the signal plus background models (solid red curve),
             and the background model alone (dotted blue curve),
             for the categories targeting \tuH FCNC interactions (\cmsLeft) and \tcH FCNC interactions (\cmsRight).
             The signal model is normalized to the best-fit value.
			 The green and yellow bands give the $\pm$1 and $\pm$2 standard deviation uncertainties in the background model (dotted blue curve). 
			 The background model includes \Hgg events from SM processes.
             Events are weighted by the $ S / (S + B) $ of their respective categories.
             The lower panels show the same information, but with the background component subtracted.
    }
    \label{fig:mgg_dist}
\end{figure}

The \mgg distributions for events entering the analysis are shown in Fig.~\ref{fig:mgg_dist}, with events weighted by the quantity $ S / (S + B)$ of their respective category.
Here $S$\,($B$) is defined as the number of signal (background) events in the range $\mH \pm \sigma_{\text{eff}}$, where $\sigma_{\text{eff}}$ is the effective width of the signal model, defined as half of the range that contains 68\% of the total number of events. 
The number of signal events $S$ is normalized to the expected 95\% \CL upper limit on \brHqt.

\begin{figure} [htb!]
    \centering
    \includegraphics[width=0.48\textwidth]{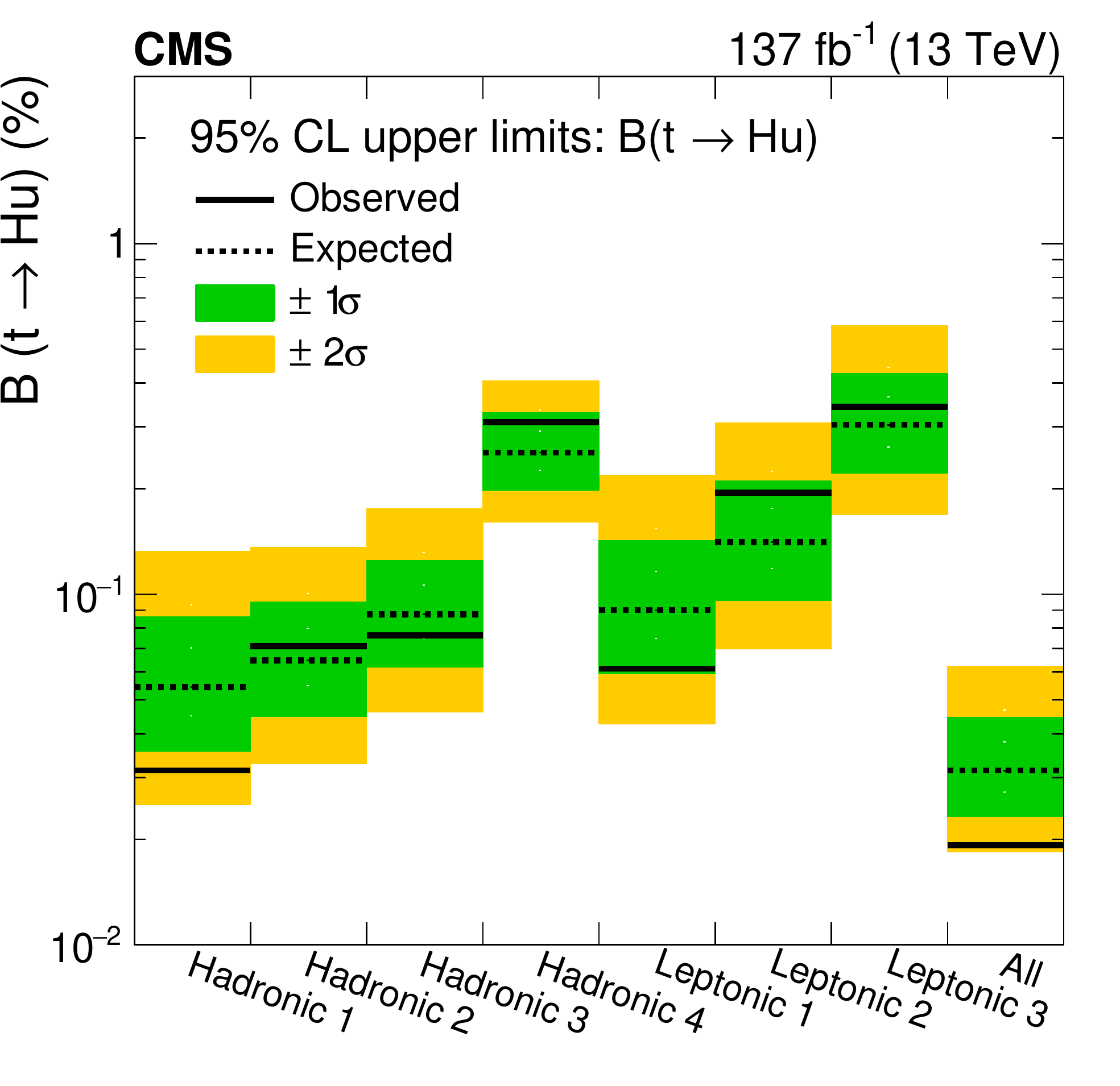}
    \includegraphics[width=0.48\textwidth]{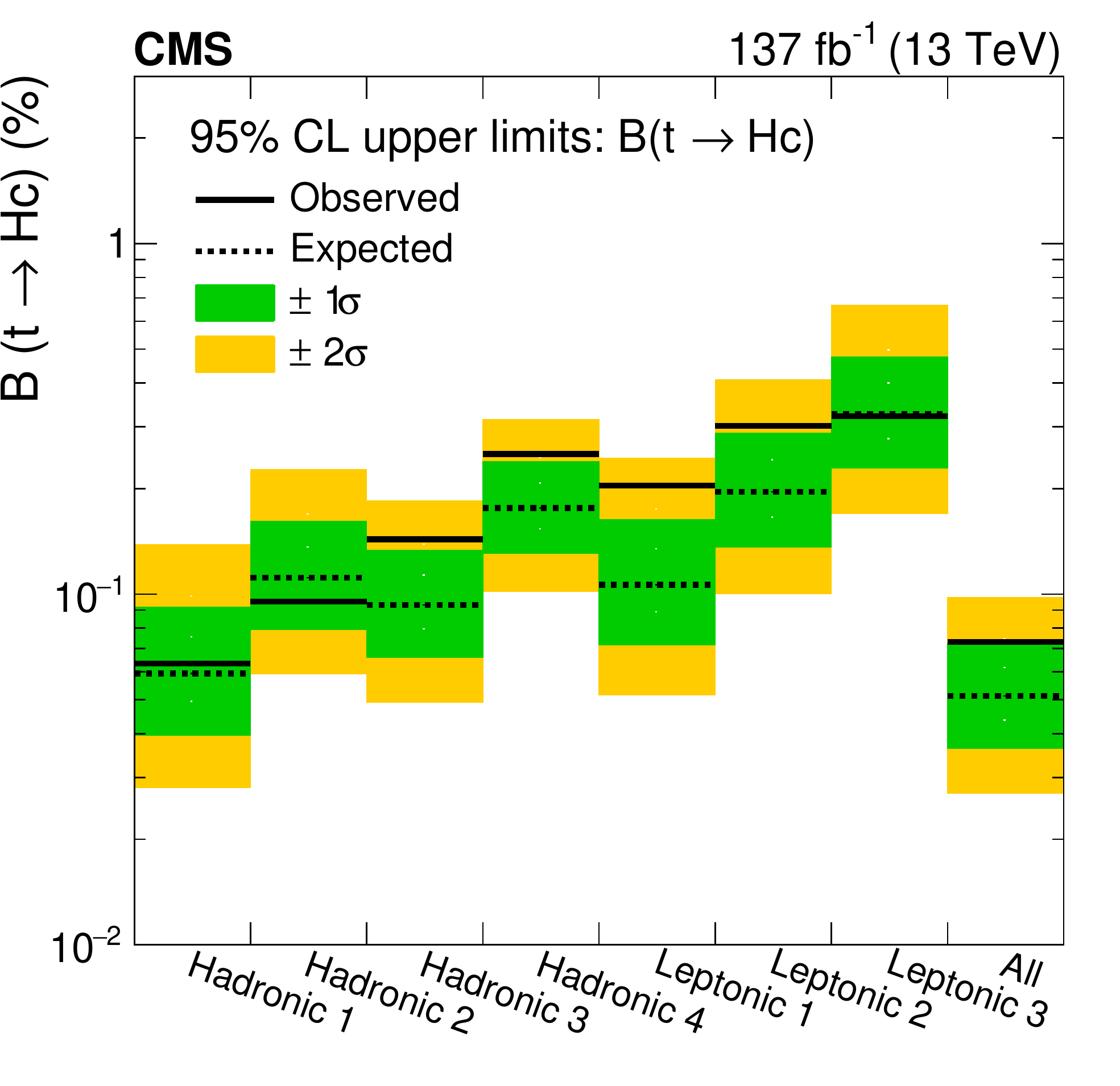}
    \caption{The observed (solid line) and expected (dotted line) 95\% \CL upper limits on \brHut (\cmsLeft) and \brHct (\cmsRight)
    for each of the hadronic and leptonic categories, defined as described in the text.
    The last bin gives the overall combined upper limit.
    The $\pm$1 and $\pm$2 standard deviation variations on the expected limit are given by the green and yellow bands, respectively.}
    \label{fig:limits}
\end{figure}

The observed and expected 95\% \CL upper limits are shown in Fig.~\ref{fig:limits} for \brHut (\cmsLeft) and \brHct (\cmsRight),
for each of the seven categories and combined.
The observed (expected) 95\% \CL upper limits on \brHut and \brHct are
$0.019\,(0.031)\%$ and $0.073\,(0.051)\%$, respectively.
The corresponding observed (expected) 95\% \CL upper limits on $\abs{\kappa_{\PH\PQu\PQt}}$ and $\abs{\kappa_{\PH\PQc\PQt}}$, derived with Eq.~(\ref{eqn:coupling_vs_bf}), are $0.037~(0.047)$ and $0.071~(0.060)$, respectively.

In summary, we have presented a search for flavor-changing neutral current interactions of the top quark (\PQt) and Higgs boson (\PH) in proton-proton collisions at a center-of-mass energy of $13$\TeV.
The processes considered include both the associated production of a single top quark with a Higgs boson via an up or charm quark, 
and the decay of a top quark to a Higgs boson and an up or charm quark in \ttbar production.
No significant excess above the background prediction is observed.
The observed (expected) 95\% confidence level upper limits on \brHut and \brHct
of $0.019\,(0.031)\%$ and $0.073\,(0.051)\%$, respectively, are the most stringent experimental limits published to date.

\begin{acknowledgments}
    We congratulate our colleagues in the CERN accelerator departments for the excellent performance of the LHC and thank the technical and administrative staffs at CERN and at other CMS institutes for their contributions to the success of the CMS effort. In addition, we gratefully acknowledge the computing centers and personnel of the Worldwide LHC Computing Grid and other centers for delivering so effectively the computing infrastructure essential to our analyses. Finally, we acknowledge the enduring support for the construction and operation of the LHC, the CMS detector, and the supporting computing infrastructure provided by the following funding agencies: BMBWF and FWF (Austria); FNRS and FWO (Belgium); CNPq, CAPES, FAPERJ, FAPERGS, and FAPESP (Brazil); MES and BNSF (Bulgaria); CERN; CAS, MoST, and NSFC (China); MINCIENCIAS (Colombia); MSES and CSF (Croatia); RIF (Cyprus); SENESCYT (Ecuador); MoER, ERC PUT and ERDF (Estonia); Academy of Finland, MEC, and HIP (Finland); CEA and CNRS/IN2P3 (France); BMBF, DFG, and HGF (Germany); GSRI (Greece); NKFIA (Hungary); DAE and DST (India); IPM (Iran); SFI (Ireland); INFN (Italy); MSIP and NRF (Republic of Korea); MES (Latvia); LAS (Lithuania); MOE and UM (Malaysia); BUAP, CINVESTAV, CONACYT, LNS, SEP, and UASLP-FAI (Mexico); MOS (Montenegro); MBIE (New Zealand); PAEC (Pakistan); MSHE and NSC (Poland); FCT (Portugal); JINR (Dubna); MON, RosAtom, RAS, RFBR, and NRC KI (Russia); MESTD (Serbia); SEIDI, CPAN, PCTI, and FEDER (Spain); MOSTR (Sri Lanka); Swiss Funding Agencies (Switzerland); MST (Taipei); ThEPCenter, IPST, STAR, and NSTDA (Thailand); TUBITAK and TAEK (Turkey); NASU (Ukraine); STFC (United Kingdom); DOE and NSF (USA).
\end{acknowledgments}

\bibliography{auto_generated}
\cleardoublepage \appendix\section{The CMS Collaboration \label{app:collab}}\begin{sloppypar}\hyphenpenalty=5000\widowpenalty=500\clubpenalty=5000\input{TOP-20-007-authorlist.tex}\end{sloppypar}
\end{document}

%% file: TOP-20-007-authorlist.tex
\cmsinstitute{Yerevan~Physics~Institute, Yerevan, Armenia}
A.~Tumasyan
\cmsinstitute{Institut~f\"{u}r~Hochenergiephysik, Vienna, Austria}
W.~Adam\cmsorcid{0000-0001-9099-4341}, J.W.~Andrejkovic, T.~Bergauer\cmsorcid{0000-0002-5786-0293}, S.~Chatterjee\cmsorcid{0000-0003-2660-0349}, M.~Dragicevic\cmsorcid{0000-0003-1967-6783}, A.~Escalante~Del~Valle\cmsorcid{0000-0002-9702-6359}, R.~Fr\"{u}hwirth\cmsAuthorMark{1}, M.~Jeitler\cmsAuthorMark{1}\cmsorcid{0000-0002-5141-9560}, N.~Krammer, L.~Lechner\cmsorcid{0000-0002-3065-1141}, D.~Liko, I.~Mikulec, P.~Paulitsch, F.M.~Pitters, J.~Schieck\cmsAuthorMark{1}\cmsorcid{0000-0002-1058-8093}, R.~Sch\"{o}fbeck\cmsorcid{0000-0002-2332-8784}, D.~Schwarz, S.~Templ\cmsorcid{0000-0003-3137-5692}, W.~Waltenberger\cmsorcid{0000-0002-6215-7228}, C.-E.~Wulz\cmsAuthorMark{1}\cmsorcid{0000-0001-9226-5812}
% ...
\vskip\cmsinstskip
% ...
1:  Also at TU~Wien, Wien, Austria\\
% ...


%% file: TOP-20-007_temp.bbl
\providecommand{\href}[2]{#2}\begingroup\raggedright\begin{thebibliography}{10}%
\makeatletter
\providecommand{\hrefCMSnoop }[0]{\@secondoftwo}%
\makeatother
\providecommand{\doi}{\texttt{doi:}\begingroup \urlstyle{tt}\Url}

\bibitem{Glashow:1970gm}
\hrefCMSnoop {}{S.~L. Glashow, J.~Iliopoulos, and L.~Maiani, ``Weak
  interactions with lepton-hadron symmetry'',} \textit{ Phys. Rev. D} \textbf{
  2} (1970) 1285,
  \href{http://dx.doi.org/10.1103/PhysRevD.2.1285}{\doi{10.1103/PhysRevD.2.1285}}.

\bibitem{Kobayashi:1973fv}
\hrefCMSnoop {}{M.~Kobayashi and T.~Maskawa, ``{CP} violation in the
  renormalizable theory of weak interaction'',} \textit{ Prog. Theor. Phys.}
  \textbf{ 49} (1973) 652,
  \href{http://dx.doi.org/10.1143/PTP.49.652}{\doi{10.1143/PTP.49.652}}.

\bibitem{Eilam:1990zc}
\hrefCMSnoop {}{G.~Eilam, J.~L. Hewett, and A.~Soni, ``Rare decays of the top
  quark in the standard and two {Higgs} doublet models'',} \textit{ Phys. Rev.
  D} \textbf{ 44} (1991) 1473,
  \href{http://dx.doi.org/10.1103/PhysRevD.44.1473}{\doi{10.1103/PhysRevD.44.1473}}.
  [Erratum: \texttt{doi:10.1103/PhysRevD.59.039901}].

\bibitem{Mele_1998}
\hrefCMSnoop {}{B.~Mele, S.~Petrarca, and A.~Soddu, ``A new evaluation of the
  decay width in the standard model'',} \textit{ Phys. Lett. B} \textbf{ 435}
  (1998) 401,
  \href{http://dx.doi.org/10.1016/s0370-2693(98)00822-3}{\doi{10.1016/s0370-2693(98)00822-3}}.

\bibitem{AguilarSaavedra:2004wm}
\href {https://www.actaphys.uj.edu.pl/R/35/11/2695/pdf}{J.~A. Aguilar-Saavedra,
  ``Top flavor-changing neutral interactions: theoretical expectations and
  experimental detection'',} \textit{ Acta Phys. Polon. B} \textbf{ 35} (2004)
  2695,
  \href{http://www.arXiv.org/abs/hep-ph/0409342}{\texttt{arXiv:hep-ph/0409342}}.

\bibitem{Zhang_2013}
\hrefCMSnoop {}{C.~Zhang and F.~Maltoni, ``Top-quark decay into {Higgs} boson
  and a light quark at next-to-leading order in {QCD}'',} \textit{ Phys. Rev.
  D} \textbf{ 88} (2013) 054005,
  \href{http://dx.doi.org/10.1103/physrevd.88.054005}{\doi{10.1103/physrevd.88.054005}},
  \href{http://www.arXiv.org/abs/1305.7386}{\texttt{arXiv:1305.7386}}.

\bibitem{Aaboud:2018oqm}
\hrefCMSnoop {}{{ATLAS Collaboration}, ``Search for top-quark decays {$\PQt \to
  \PH\PQq$} with {36 fb$^{-1}$} of {$\Pp\Pp$} collision data at {$\sqrt{s}=13$
  TeV} with the {ATLAS} detector'',} \textit{ JHEP} \textbf{ 05} (2019) 123,
  \href{http://dx.doi.org/10.1007/JHEP05(2019)123}{\doi{10.1007/JHEP05(2019)123}},
  \href{http://www.arXiv.org/abs/1812.11568}{\texttt{arXiv:1812.11568}}.

\bibitem{Azatov_2009}
\hrefCMSnoop {}{A.~Azatov, M.~Toharia, and L.~Zhu, ``Higgs mediated flavor
  changing neutral currents in warped extra dimensions'',} \textit{ Phys. Rev.
  D} \textbf{ 80} (2009) 035016,
  \href{http://dx.doi.org/10.1103/physrevd.80.035016}{\doi{10.1103/physrevd.80.035016}},
  \href{http://www.arXiv.org/abs/0906.1990}{\texttt{arXiv:0906.1990}}.

\bibitem{Azatov_2014}
\hrefCMSnoop {}{A.~Azatov, G.~Panico, G.~Perez, and Y.~Soreq, ``On the flavor
  structure of natural composite {Higgs} models \& top flavor violation'',}
  \textit{ JHEP} \textbf{ 12} (2014) 082,
  \href{http://dx.doi.org/10.1007/jhep12(2014)082}{\doi{10.1007/jhep12(2014)082}},
  \href{http://www.arXiv.org/abs/1408.4525}{\texttt{arXiv:1408.4525}}.

\bibitem{Bejar:2000ub}
\hrefCMSnoop {}{S.~Bejar, J.~Guasch, and J.~Sol{\`{a}}, ``Loop induced flavor
  changing neutral decays of the top quark in a general two {Higgs} doublet
  model'',} \textit{ Nucl. Phys. B} \textbf{ 600} (2001) 21,
  \href{http://dx.doi.org/10.1016/S0550-3213(01)00044-X}{\doi{10.1016/S0550-3213(01)00044-X}},
  \href{http://www.arXiv.org/abs/hep-ph/0011091}{\texttt{arXiv:hep-ph/0011091}}.

\bibitem{Guasch_1999}
\hrefCMSnoop {}{J.~Guasch and J.~Sol{\`{a}}, ``{FCNC} top quark decays in the
  {MSSM}: a door to {SUSY} physics in high luminosity colliders?'',} \textit{
  Nucl. Phys. B} \textbf{ 562} (1999) 3,
  \href{http://dx.doi.org/10.1016/s0550-3213(99)00579-9}{\doi{10.1016/s0550-3213(99)00579-9}}.

\bibitem{Cao_2007}
J.~J. Cao\hrefCMSnoop {}{ {et~al.}, ``Supersymmetry-induced flavor-changing
  neutral-current top-quark processes at the {CERN} {Large Hadron Collider}'',}
  \textit{ Phys. Rev. D} \textbf{ 75} (2007) 075021,
  \href{http://dx.doi.org/10.1103/physrevd.75.075021}{\doi{10.1103/physrevd.75.075021}},
  \href{http://www.arXiv.org/abs/hep-ph/0702264}{\texttt{arXiv:hep-ph/0702264}}.

\bibitem{Cao_2014}
J.~Cao\hrefCMSnoop {}{ {et~al.}, ``{SUSY} induced top quark {FCNC} decay {$\PQt
  \to \PQc\PH$} after {Run I} of {LHC}'',} \textit{ Eur. Phys. J. C} \textbf{
  74} (2014) 3058,
  \href{http://dx.doi.org/10.1140/epjc/s10052-014-3058-1}{\doi{10.1140/epjc/s10052-014-3058-1}},
  \href{http://www.arXiv.org/abs/1404.1241}{\texttt{arXiv:1404.1241}}.

\bibitem{Eilam_2001}
G.~Eilam\hrefCMSnoop {}{ {et~al.}, ``Top-quark rare decay {$\PQt \to \PQc\PH$}
  in {R-parity-violating SUSY}'',} \textit{ Phys. Lett. B} \textbf{ 510} (2001)
  227,
  \href{http://dx.doi.org/10.1016/s0370-2693(01)00598-6}{\doi{10.1016/s0370-2693(01)00598-6}},
  \href{http://www.arXiv.org/abs/hep-ph/0102037}{\texttt{arXiv:hep-ph/0102037}}.

\bibitem{AguilarSaavedra:2002kr}
\hrefCMSnoop {}{J.~A. Aguilar-Saavedra, ``Effects of mixing with quark
  singlets'',} \textit{ Phys. Rev. D} \textbf{ 67} (2003) 035003,
  \href{http://dx.doi.org/10.1103/PhysRevD.69.099901}{\doi{10.1103/PhysRevD.69.099901}},
  \href{http://www.arXiv.org/abs/hep-ph/0210112}{\texttt{arXiv:hep-ph/0210112}}.
  [Erratum: \texttt{doi:10.1103/PhysRevD.67.035003}].

\bibitem{Hou:1991un}
\hrefCMSnoop {}{W.-S. Hou, ``{Tree level t \ensuremath{\to} c h or h
  \ensuremath{\to} t c decays}'',} \textit{ Phys. Lett. B} \textbf{ 296} (1992)
  179,
  \href{http://dx.doi.org/10.1016/0370-2693(92)90823-M}{\doi{10.1016/0370-2693(92)90823-M}}.

\bibitem{Chen:2013qta}
\hrefCMSnoop {}{K.-F. Chen, W.-S. Hou, C.~Kao, and M.~Kohda, ``{When the Higgs
  meets the top: Search for $t \to ch^0$ at the LHC}'',} \textit{ Phys. Lett.
  B} \textbf{ 725} (2013) 378,
  \href{http://dx.doi.org/10.1016/j.physletb.2013.07.060}{\doi{10.1016/j.physletb.2013.07.060}},
  \href{http://www.arXiv.org/abs/1304.8037}{\texttt{arXiv:1304.8037}}.

\bibitem{Baum:2008qm}
\hrefCMSnoop {}{I.~Baum, G.~Eilam, and S.~Bar-Shalom, ``{Scalar flavor changing
  neutral currents and rare top quark decays in a two H iggs doublet model 'for
  the top quark'}'',} \textit{ Phys. Rev. D} \textbf{ 77} (2008) 113008,
  \href{http://dx.doi.org/10.1103/PhysRevD.77.113008}{\doi{10.1103/PhysRevD.77.113008}},
  \href{http://www.arXiv.org/abs/0802.2622}{\texttt{arXiv:0802.2622}}.

\bibitem{Kao:2011aa}
\hrefCMSnoop {}{C.~Kao, H.-Y. Cheng, W.-S. Hou, and J.~Sayre, ``{Top decays
  with flavor changing neutral Higgs interactions at the LHC}'',} \textit{
  Phys. Lett. B} \textbf{ 716} (2012) 225,
  \href{http://dx.doi.org/10.1016/j.physletb.2012.08.032}{\doi{10.1016/j.physletb.2012.08.032}},
  \href{http://www.arXiv.org/abs/1112.1707}{\texttt{arXiv:1112.1707}}.

\bibitem{Altunkaynak:2015twa}
B.~Altunkaynak\hrefCMSnoop {}{ {et~al.}, ``{Flavor changing heavy Higgs
  interactions at the LHC}'',} \textit{ Phys. Lett. B} \textbf{ 751} (2015)
  135,
  \href{http://dx.doi.org/10.1016/j.physletb.2015.10.024}{\doi{10.1016/j.physletb.2015.10.024}},
  \href{http://www.arXiv.org/abs/1506.00651}{\texttt{arXiv:1506.00651}}.

\bibitem{Aaboud:2017mfd}
\hrefCMSnoop {}{{ATLAS Collaboration}, ``Search for top quark decays {$\PQt \to
  \PQq\PH$}, with {$\PH \to \PGg\PGg$}, in {$\sqrt{s}=13$ TeV $\Pp\Pp$}
  collisions using the {ATLAS} detector'',} \textit{ JHEP} \textbf{ 10} (2017)
  129,
  \href{http://dx.doi.org/10.1007/JHEP10(2017)129}{\doi{10.1007/JHEP10(2017)129}},
  \href{http://www.arXiv.org/abs/1707.01404}{\texttt{arXiv:1707.01404}}.

\bibitem{Aaboud:2018pob}
\hrefCMSnoop {}{{ATLAS Collaboration}, ``Search for flavor-changing neutral
  currents in top quark decays {$\PQt \to \PH\PQc$} and {$\PQt \to \PH\PQu$} in
  multilepton final states in proton-proton collisions at {$\sqrt{s}= 13$ TeV}
  with the {ATLAS} detector'',} \textit{ Phys. Rev. D} \textbf{ 98} (2018)
  032002,
  \href{http://dx.doi.org/10.1103/PhysRevD.98.032002}{\doi{10.1103/PhysRevD.98.032002}},
  \href{http://www.arXiv.org/abs/1805.03483}{\texttt{arXiv:1805.03483}}.

\bibitem{Sirunyan:2017uae}
\hrefCMSnoop {}{{CMS Collaboration}, ``Search for the flavor-changing neutral
  current interactions of the top quark and the {Higgs} boson which decays into
  a pair of b quarks at {$\sqrt{s}=$ 13 TeV}'',} \textit{ JHEP} \textbf{ 06}
  (2018) 102,
  \href{http://dx.doi.org/10.1007/JHEP06(2018)102}{\doi{10.1007/JHEP06(2018)102}},
  \href{http://www.arXiv.org/abs/1712.02399}{\texttt{arXiv:1712.02399}}.

\bibitem{hepdata}
\hrefCMSnoop {}{}{HEPD}ata record for this analysis, 2021.
\newblock
  \href{http://dx.doi.org/10.17182/hepdata.105999}{\doi{10.17182/hepdata.105999}}.

\bibitem{Chatrchyan:2008zzk}
\hrefCMSnoop {}{{CMS Collaboration}, ``The {CMS} experiment at the {CERN}
  {LHC}'',} \textit{ JINST} \textbf{ 3} (2008) S08004,
\href{http://dx.doi.org/10.1088/1748-0221/3/08/S08004}{\doi{10.1088/1748-0221/3/08/S08004}}.

\bibitem{Alwall:2014hca}
J.~Alwall\hrefCMSnoop {}{ {et~al.}, ``The automated computation of tree-level
  and next-to-leading order differential cross sections, and their matching to
  parton shower simulations'',} \textit{ JHEP} \textbf{ 07} (2014) 079,
  \href{http://dx.doi.org/10.1007/JHEP07(2014)079}{\doi{10.1007/JHEP07(2014)079}},
\href{http://www.arXiv.org/abs/1405.0301}{\texttt{arXiv:1405.0301}}.

\bibitem{Sjostrand:2014zea}
T.~Sj{\"o}strand\hrefCMSnoop {}{ {et~al.}, ``An introduction to {PYTHIA}
  8.2'',} \textit{ Comput. Phys. Commun.} \textbf{ 191} (2015) 159,
  \href{http://dx.doi.org/10.1016/j.cpc.2015.01.024}{\doi{10.1016/j.cpc.2015.01.024}},
\href{http://www.arXiv.org/abs/1410.3012}{\texttt{arXiv:1410.3012}}.

\bibitem{Khachatryan:2015pea}
\hrefCMSnoop {}{{CMS Collaboration}, ``Event generator tunes obtained from
  underlying event and multiparton scattering measurements'',} \textit{ Eur.
  Phys. J. C} \textbf{ 76} (2016) 155,
  \href{http://dx.doi.org/10.1140/epjc/s10052-016-3988-x}{\doi{10.1140/epjc/s10052-016-3988-x}},
\href{http://www.arXiv.org/abs/1512.00815}{\texttt{arXiv:1512.00815}}.

\bibitem{Sirunyan:2019dfx}
\hrefCMSnoop {}{{CMS Collaboration}, ``Extraction and validation of a new set
  of {CMS} {PYTHIA8} tunes from underlying-event measurements'',} \textit{ Eur.
  Phys. J. C} \textbf{ 80} (2020) 4,
  \href{http://dx.doi.org/10.1140/epjc/s10052-019-7499-4}{\doi{10.1140/epjc/s10052-019-7499-4}},
  \href{http://www.arXiv.org/abs/1903.12179}{\texttt{arXiv:1903.12179}}.

\bibitem{NNPDF3}
\hrefCMSnoop {}{{NNPDF} Collaboration, ``Parton distributions for the {LHC Run
  II}'',} \textit{ JHEP} \textbf{ 04} (2015) 040,
  \href{http://dx.doi.org/10.1007/JHEP04(2015)040}{\doi{10.1007/JHEP04(2015)040}},
\href{http://www.arXiv.org/abs/1410.8849}{\texttt{arXiv:1410.8849}}.

\bibitem{NNPDF31}
\hrefCMSnoop {}{{NNPDF} Collaboration, ``{Parton distributions from
  high-precision collider data}'',} \textit{ Eur. Phys. J. C} \textbf{ 77}
  (2017) 663,
  \href{http://dx.doi.org/10.1140/epjc/s10052-017-5199-5}{\doi{10.1140/epjc/s10052-017-5199-5}},
  \href{http://www.arXiv.org/abs/1706.00428}{\texttt{arXiv:1706.00428}}.

\bibitem{Buchmuller:1985jz}
\hrefCMSnoop {}{W.~Buchmuller and D.~Wyler, ``Effective {Lagrangian} analysis
  of new interactions and flavor conservation'',} \textit{ Nucl. Phys. B}
  \textbf{ 268} (1986) 621,
  \href{http://dx.doi.org/10.1016/0550-3213(86)90262-2}{\doi{10.1016/0550-3213(86)90262-2}}.

\bibitem{Grzadkowski:2010es}
\hrefCMSnoop {}{B.~Grzadkowski, M.~Iskrzynski, M.~Misiak, and J.~Rosiek,
  ``Dimension-six terms in the {Standard Model Lagrangian}'',} \textit{ JHEP}
  \textbf{ 10} (2010) 085,
  \href{http://dx.doi.org/10.1007/JHEP10(2010)085}{\doi{10.1007/JHEP10(2010)085}},
  \href{http://www.arXiv.org/abs/1008.4884}{\texttt{arXiv:1008.4884}}.

\bibitem{Alloul:2013bka}
A.~Alloul\hrefCMSnoop {}{ {et~al.}, ``{FeynRules 2.0 - A complete toolbox for
  tree-level phenomenology}'',} \textit{ Comput. Phys. Commun.} \textbf{ 185}
  (2014) 2250,
  \href{http://dx.doi.org/10.1016/j.cpc.2014.04.012}{\doi{10.1016/j.cpc.2014.04.012}},
\href{http://www.arXiv.org/abs/1310.1921}{\texttt{arXiv:1310.1921}}.

\bibitem{Degrande:2011ua}
C.~Degrande\hrefCMSnoop {}{ {et~al.}, ``{UFO --- The Universal FeynRules
  Output}'',} \textit{ Comput. Phys. Commun.} \textbf{ 183} (2012) 1201,
  \href{http://dx.doi.org/10.1016/j.cpc.2012.01.022}{\doi{10.1016/j.cpc.2012.01.022}},
\href{http://www.arXiv.org/abs/1108.2040}{\texttt{arXiv:1108.2040}}.

\bibitem{Alwall:2007fs}
J.~Alwall\hrefCMSnoop {}{ {et~al.}, ``Comparative study of various algorithms
  for the merging of parton showers and matrix elements in hadronic
  collisions'',} \textit{ Eur. Phys. J. C} \textbf{ 53} (2008) 473,
  \href{http://dx.doi.org/10.1140/epjc/s10052-007-0490-5}{\doi{10.1140/epjc/s10052-007-0490-5}},
\href{http://www.arXiv.org/abs/0706.2569}{\texttt{arXiv:0706.2569}}.

\bibitem{Czakon:2011xx}
\hrefCMSnoop {}{M.~Czakon and A.~Mitov, ``Top++: A program for the calculation
  of the top-pair cross-section at hadron colliders'',} \textit{ Comput. Phys.
  Commun.} \textbf{ 185} (2014) 2930,
  \href{http://dx.doi.org/10.1016/j.cpc.2014.06.021}{\doi{10.1016/j.cpc.2014.06.021}},
\href{http://www.arXiv.org/abs/1112.5675}{\texttt{arXiv:1112.5675}}.

\bibitem{Gao:2012ja}
\hrefCMSnoop {}{J.~Gao, C.~S. Li, and H.~X. Zhu, ``Top quark decay at
  next-to-next-to leading order in {QCD}'',} \textit{ Phys. Rev. Lett.}
  \textbf{ 110} (2013) 042001,
  \href{http://dx.doi.org/10.1103/PhysRevLett.110.042001}{\doi{10.1103/PhysRevLett.110.042001}},
  \href{http://www.arXiv.org/abs/1210.2808}{\texttt{arXiv:1210.2808}}.

\bibitem{Kidonakis:2012rm}
\hrefCMSnoop {}{N.~Kidonakis, ``{NNLL threshold resummation for top-pair and
  single-top production}'',} \textit{ Phys. Part. Nucl.} \textbf{ 45} (2014)
  714,
  \href{http://dx.doi.org/10.1134/S1063779614040091}{\doi{10.1134/S1063779614040091}},
  \href{http://www.arXiv.org/abs/1210.7813}{\texttt{arXiv:1210.7813}}.

\bibitem{Czakon:2015owf}
\hrefCMSnoop {}{M.~Czakon, D.~Heymes, and A.~Mitov, ``{High-precision
  differential predictions for top-quark pairs at the LHC}'',} \textit{ Phys.
  Rev. Lett.} \textbf{ 116} (2016) 082003,
  \href{http://dx.doi.org/10.1103/PhysRevLett.116.082003}{\doi{10.1103/PhysRevLett.116.082003}},
  \href{http://www.arXiv.org/abs/1511.00549}{\texttt{arXiv:1511.00549}}.

\bibitem{Czakon:2017wor}
M.~Czakon\hrefCMSnoop {}{ {et~al.}, ``{Top-pair production at the LHC through
  NNLO QCD and NLO EW}'',} \textit{ JHEP} \textbf{ 10} (2017) 186,
  \href{http://dx.doi.org/10.1007/JHEP10(2017)186}{\doi{10.1007/JHEP10(2017)186}},
  \href{http://www.arXiv.org/abs/1705.04105}{\texttt{arXiv:1705.04105}}.

\bibitem{Catani:2019hip}
S.~Catani\hrefCMSnoop {}{ {et~al.}, ``{Top-quark pair production at the LHC:
  fully differential QCD predictions at NNLO}'',} \textit{ JHEP} \textbf{ 07}
  (2019) 100,
  \href{http://dx.doi.org/10.1007/JHEP07(2019)100}{\doi{10.1007/JHEP07(2019)100}},
  \href{http://www.arXiv.org/abs/1906.06535}{\texttt{arXiv:1906.06535}}.

\bibitem{deFlorian:2016spz}
\hrefCMSnoop {}{{LHC Higgs Cross Section Working Group}, ``Handbook of {LHC
  Higgs} cross sections: 4. deciphering the nature of the {Higgs} sector'',}
  \textit{ CERN} (2016)
  \href{http://dx.doi.org/10.23731/CYRM-2017-002}{\doi{10.23731/CYRM-2017-002}},
\href{http://www.arXiv.org/abs/1610.07922}{\texttt{arXiv:1610.07922}}.

\bibitem{Nason:2004rx}
\hrefCMSnoop {}{P.~Nason, ``A new method for combining {NLO QCD} with shower
  {Monte Carlo} algorithms'',} \textit{ JHEP} \textbf{ 11} (2004) 040,
  \href{http://dx.doi.org/10.1088/1126-6708/2004/11/040}{\doi{10.1088/1126-6708/2004/11/040}},
\href{http://www.arXiv.org/abs/hep-ph/0409146}{\texttt{arXiv:hep-ph/0409146}}.

\bibitem{Frixione:2007vw}
\hrefCMSnoop {}{S.~Frixione, P.~Nason, and C.~Oleari, ``{Matching NLO QCD
  computations with parton shower simulations: the \POWHEG method}'',} \textit{
  JHEP} \textbf{ 11} (2007) 070,
  \href{http://dx.doi.org/10.1088/1126-6708/2007/11/070}{\doi{10.1088/1126-6708/2007/11/070}},
\href{http://www.arXiv.org/abs/0709.2092}{\texttt{arXiv:0709.2092}}.

\bibitem{Alioli:2010xd}
\hrefCMSnoop {}{S.~Alioli, P.~Nason, C.~Oleari, and E.~Re, ``{A general
  framework for implementing NLO calculations in shower Monte Carlo programs:
  the \POWHEG BOX}'',} \textit{ JHEP} \textbf{ 06} (2010) 043,
  \href{http://dx.doi.org/10.1007/JHEP06(2010)043}{\doi{10.1007/JHEP06(2010)043}},
\href{http://www.arXiv.org/abs/1002.2581}{\texttt{arXiv:1002.2581}}.

\bibitem{Hartanto:2015uka}
\hrefCMSnoop {}{H.~B. Hartanto, B.~Jager, L.~Reina, and D.~Wackeroth, ``{Higgs
  boson production in association with top quarks in the \POWHEG BOX}'',}
  \textit{ Phys. Rev. D} \textbf{ 91} (2015) 094003,
  \href{http://dx.doi.org/10.1103/PhysRevD.91.094003}{\doi{10.1103/PhysRevD.91.094003}},
\href{http://www.arXiv.org/abs/1501.04498}{\texttt{arXiv:1501.04498}}.

\bibitem{10.21468/SciPostPhys.7.3.034}
E.~Bothmann\hrefCMSnoop {}{ {et~al.}, ``Event generation with {SHERPA} 2.2'',}
  \textit{ SciPost Phys.} \textbf{ 7} (2019) 34,
  \href{http://dx.doi.org/10.21468/SciPostPhys.7.3.034}{\doi{10.21468/SciPostPhys.7.3.034}},
  \href{http://www.arXiv.org/abs/1905.09127}{\texttt{arXiv:1905.09127}}.

\bibitem{Agostinelli:2002hh}
\hrefCMSnoop {}{{\GEANTfour} Collaboration, ``{\GEANTfour} --- a simulation
  toolkit'',} \textit{ Nucl. Instrum. Meth. A} \textbf{ 506} (2003) 250,
\href{http://dx.doi.org/10.1016/S0168-9002(03)01368-8}{\doi{10.1016/S0168-9002(03)01368-8}}.

\bibitem{Khachatryan:2016bia}
\hrefCMSnoop {}{{CMS Collaboration}, ``The {CMS} trigger system'',} \textit{
  JINST} \textbf{ 12} (2017) P01020,
  \href{http://dx.doi.org/10.1088/1748-0221/12/01/P01020}{\doi{10.1088/1748-0221/12/01/P01020}},
\href{http://www.arXiv.org/abs/1609.02366}{\texttt{arXiv:1609.02366}}.

\bibitem{Sirunyan:2018ouh}
\hrefCMSnoop {}{{CMS Collaboration}, ``Measurements of {Higgs} boson properties
  in the diphoton decay channel in proton-proton collisions at {$\sqrt{s} =$ 13
  TeV}'',} \textit{ JHEP} \textbf{ 11} (2018) 185,
  \href{http://dx.doi.org/10.1007/JHEP11(2018)185}{\doi{10.1007/JHEP11(2018)185}},
\href{http://www.arXiv.org/abs/1804.02716}{\texttt{arXiv:1804.02716}}.

\bibitem{CMS:2011aa}
\hrefCMSnoop {}{{CMS Collaboration}, ``Measurement of the inclusive {$\PW$} and
  {$\PZ$} production cross sections in pp collisions at $\sqrt{s}=7$ {TeV}'',}
  \textit{ JHEP} \textbf{ 10} (2011) 132,
  \href{http://dx.doi.org/10.1007/JHEP10(2011)132}{\doi{10.1007/JHEP10(2011)132}},
\href{http://www.arXiv.org/abs/1107.4789}{\texttt{arXiv:1107.4789}}.

\bibitem{CMS-PRF-14-001}
\hrefCMSnoop {}{{CMS Collaboration}, ``Particle-flow reconstruction and global
  event description with the {CMS} detector'',} \textit{ JINST} \textbf{ 12}
  (2017) P10003,
  \href{http://dx.doi.org/10.1088/1748-0221/12/10/P10003}{\doi{10.1088/1748-0221/12/10/P10003}},
\href{http://www.arXiv.org/abs/1706.04965}{\texttt{arXiv:1706.04965}}.

\bibitem{Cacciari:2008gp}
\hrefCMSnoop {}{M.~Cacciari, G.~P. Salam, and G.~Soyez, ``The anti-\kt jet
  clustering algorithm'',} \textit{ JHEP} \textbf{ 04} (2008) 063,
  \href{http://dx.doi.org/10.1088/1126-6708/2008/04/063}{\doi{10.1088/1126-6708/2008/04/063}},
\href{http://www.arXiv.org/abs/0802.1189}{\texttt{arXiv:0802.1189}}.

\bibitem{Cacciari:2011ma}
\hrefCMSnoop {}{M.~Cacciari, G.~P. Salam, and G.~Soyez, ``{FastJet} user
  manual'',} \textit{ Eur. Phys. J. C} \textbf{ 72} (2012) 1896,
  \href{http://dx.doi.org/10.1140/epjc/s10052-012-1896-2}{\doi{10.1140/epjc/s10052-012-1896-2}},
\href{http://www.arXiv.org/abs/1111.6097}{\texttt{arXiv:1111.6097}}.

\bibitem{Sirunyan:2019kia}
\hrefCMSnoop {}{{CMS Collaboration}, ``Performance of missing transverse
  momentum reconstruction in proton-proton collisions at $\sqrt{s} = 13$\,{TeV}
  using the {CMS} detector'',} \textit{ JINST} \textbf{ 14} (2019) P07004,
  \href{http://dx.doi.org/10.1088/1748-0221/14/07/P07004}{\doi{10.1088/1748-0221/14/07/P07004}},
\href{http://www.arXiv.org/abs/1903.06078}{\texttt{arXiv:1903.06078}}.

\bibitem{Sirunyan:2017ezt}
\hrefCMSnoop {}{{CMS Collaboration}, ``Identification of heavy-flavor jets with
  the {CMS} detector in {$\Pp\Pp$} collisions at 13 {TeV}'',} \textit{ JINST}
  \textbf{ 13} (2018) P05011,
  \href{http://dx.doi.org/10.1088/1748-0221/13/05/P05011}{\doi{10.1088/1748-0221/13/05/P05011}},
\href{http://www.arXiv.org/abs/1712.07158}{\texttt{arXiv:1712.07158}}.

\bibitem{Bols:2020bkb}
E.~Bols\hrefCMSnoop {}{ {et~al.}, ``Jet flavour classification using
  {DeepJet}'',} \textit{ JINST} \textbf{ 15} (2020) P12012,
  \href{http://dx.doi.org/10.1088/1748-0221/15/12/P12012}{\doi{10.1088/1748-0221/15/12/P12012}},
  \href{http://www.arXiv.org/abs/2008.10519}{\texttt{arXiv:2008.10519}}.

\bibitem{CMS-DP-2018-058}
\href {http://cds.cern.ch/record/2646773}{{CMS Collaboration}, ``Performance of
  the {DeepJet} b tagging algorithm using {41.9 fb$^{-1}$} of data from
  proton-proton collisions at 13 {TeV} with {Phase-1} {CMS} detector'',} CMS
  Detector Performance Note CMS-DP-2018-058, 2018.

\bibitem{DBLP:journals/corr/abs-1211-6581}
\hrefCMSnoop {}{E.~Spyromitros-Xioufis, W.~Groves, G.~Tsoumakas, and
  I.~Vlahavas, ``Multi-target regression via input space expansion: treating
  targets as inputs'',} \textit{ Mach. Learn.} \textbf{ 104} (2016) 55,
  \href{http://dx.doi.org/10.1007/s10994-016-5546-z}{\doi{10.1007/s10994-016-5546-z}},
  \href{http://www.arXiv.org/abs/1211.6581}{\texttt{arXiv:1211.6581}}.

\bibitem{Sirunyan:2018fpa}
\hrefCMSnoop {}{{CMS Collaboration}, ``{Performance of the CMS muon detector
  and muon reconstruction with proton-proton collisions at $\sqrt{s}=$ 13
  TeV}'',} \textit{ JINST} \textbf{ 13} (2018) P06015,
  \href{http://dx.doi.org/10.1088/1748-0221/13/06/P06015}{\doi{10.1088/1748-0221/13/06/P06015}},
  \href{http://www.arXiv.org/abs/1804.04528}{\texttt{arXiv:1804.04528}}.

\bibitem{xgboost}
\hrefCMSnoop {}{T.~Chen and C.~Guestrin, ``{XGBoost}: A scalable tree boosting
  system'',} in \textit{ Proc. 22nd ACM SIGKDD Intern. Conf. on Knowledge
  Discovery and Data Mining}, KDD, p.~785.
\newblock ACM, New York, NY, USA, 2016.
\newblock
  \href{http://dx.doi.org/10.1145/2939672.2939785}{\doi{10.1145/2939672.2939785}}.

\bibitem{Sirunyan:2020sum}
\hrefCMSnoop {}{{CMS Collaboration}, ``Measurements of {$\ttbar\PH$} production
  and the {CP} structure of the {Yukawa} interaction between the {Higgs} boson
  and top quark in the diphoton decay channel'',} \textit{ Phys. Rev. Lett.}
  \textbf{ 125} (2020) 061801,
  \href{http://dx.doi.org/10.1103/PhysRevLett.125.061801}{\doi{10.1103/PhysRevLett.125.061801}},
  \href{http://www.arXiv.org/abs/2003.10866}{\texttt{arXiv:2003.10866}}.

\bibitem{CMS:2018hfk}
\hrefCMSnoop {}{{CMS Collaboration}, ``{Evidence for the associated production
  of a single top quark and a photon in proton-proton collisions at $\sqrt{s}=$
  13 TeV}'',} \textit{ Phys. Rev. Lett.} \textbf{ 121} (2018), no.~22, 221802,
  \href{http://dx.doi.org/10.1103/PhysRevLett.121.221802}{\doi{10.1103/PhysRevLett.121.221802}},
  \href{http://www.arXiv.org/abs/1808.02913}{\texttt{arXiv:1808.02913}}.

\bibitem{tmva}
\hrefCMSnoop {}{H.~Voss, A.~H{\"o}cker, J.~Stelzer, and F.~Tegenfeldt,
  ``{TMVA}, the toolkit for multivariate data analysis with {ROOT}'',} in
  \textit{ XIth International Workshop on Advanced Computing and Analysis
  Techniques in Physics Research (ACAT)}, p.~40.
\newblock 2007.
\newblock
  \href{http://www.arXiv.org/abs/physics/0703039}{\texttt{arXiv:physics/0703039}}.
\newblock
\href{http://dx.doi.org/10.22323/1.050.0040}{\doi{10.22323/1.050.0040}}.

\bibitem{CrystalBallRef}
\href {http://www.slac.stanford.edu/cgi-wrap/getdoc/slac-r-236.pdf}{M.~J.
  Oreglia, ``A study of the reactions $\psi^\prime \to \gamma \gamma \psi$''}.
\newblock PhD thesis, Stanford University, 1980.
\newblock {SLAC} Report {SLAC-R-236}.

\bibitem{CMS:2019drq}
\hrefCMSnoop {}{{CMS Collaboration}, ``{A measurement of the Higgs boson mass
  in the diphoton decay channel}'',} \textit{ Phys. Lett. B} \textbf{ 805}
  (2020) 135425,
  \href{http://dx.doi.org/10.1016/j.physletb.2020.135425}{\doi{10.1016/j.physletb.2020.135425}},
  \href{http://www.arXiv.org/abs/2002.06398}{\texttt{arXiv:2002.06398}}.

\bibitem{Dauncey:2014xga}
\hrefCMSnoop {}{P.~D. Dauncey, M.~Kenzie, N.~Wardle, and G.~J. Davies,
  ``Handling uncertainties in background shapes'',} \textit{ JINST} \textbf{
  10} (2015) P04015,
  \href{http://dx.doi.org/10.1088/1748-0221/10/04/P04015}{\doi{10.1088/1748-0221/10/04/P04015}},
\href{http://www.arXiv.org/abs/1408.6865}{\texttt{arXiv:1408.6865}}.

\bibitem{Butterworth:2015oua}
\hrefCMSnoop {}{J.~Butterworth {et~al.}, ``{PDF4LHC recommendations for LHC Run
  II}'',} \textit{ J. Phys. G} \textbf{ 43} (2016) 023001,
  \href{http://dx.doi.org/10.1088/0954-3899/43/2/023001}{\doi{10.1088/0954-3899/43/2/023001}},
  \href{http://www.arXiv.org/abs/1510.03865}{\texttt{arXiv:1510.03865}}.

\bibitem{Martin:2009bu}
\hrefCMSnoop {}{A.~D. Martin, W.~J. Stirling, R.~S. Thorne, and G.~Watt,
  ``Uncertainties on {$\alpha_S$} in global {PDF} analyses and implications for
  predicted hadronic cross sections'',} \textit{ Eur. Phys. J. C} \textbf{ 64}
  (2009) 653,
  \href{http://dx.doi.org/10.1140/epjc/s10052-009-1164-2}{\doi{10.1140/epjc/s10052-009-1164-2}},
  \href{http://www.arXiv.org/abs/0905.3531}{\texttt{arXiv:0905.3531}}.

\bibitem{Gao:2013xoa}
J.~Gao\hrefCMSnoop {}{ {et~al.}, ``{CT10} next-to-next-to-leading order global
  analysis of {QCD}'',} \textit{ Phys. Rev. D} \textbf{ 89} (2014) 033009,
  \href{http://dx.doi.org/10.1103/PhysRevD.89.033009}{\doi{10.1103/PhysRevD.89.033009}},
  \href{http://www.arXiv.org/abs/1302.6246}{\texttt{arXiv:1302.6246}}.

\bibitem{Ball:2012cx}
\hrefCMSnoop {}{{NNPDF} Collaboration, ``{Parton distributions with LHC
  data}'',} \textit{ Nucl. Phys. B} \textbf{ 867} (2013) 244,
  \href{http://dx.doi.org/10.1016/j.nuclphysb.2012.10.003}{\doi{10.1016/j.nuclphysb.2012.10.003}},
\href{http://www.arXiv.org/abs/1207.1303}{\texttt{arXiv:1207.1303}}.

\bibitem{CMS-LUM-17-003}
\hrefCMSnoop {}{{CMS Collaboration}, ``Precision luminosity measurement in
  proton-proton collisions at $\sqrt{s} =$ 13 {TeV} in 2015 and 2016 at
  {CMS}'',} \textit{ Eur. Phys. J. C} \textbf{ 81} (2021) 800,
  \href{http://dx.doi.org/10.1140/epjc/s10052-021-09538-2}{\doi{10.1140/epjc/s10052-021-09538-2}},
  \href{http://www.arXiv.org/abs/2104.01927}{\texttt{arXiv:2104.01927}}.

\bibitem{CMS-PAS-LUM-17-004}
\href {https://cds.cern.ch/record/2621960}{{CMS Collaboration}, ``{CMS}
  luminosity measurement for the 2017 data-taking period at {$\sqrt{s} =
  13~\mathrm{TeV}$}'',} CMS Physics Analysis Summary CMS-PAS-LUM-17-004, 2018.

\bibitem{CMS-PAS-LUM-18-002}
\href {https://cds.cern.ch/record/2676164}{{CMS Collaboration}, ``{CMS}
  luminosity measurement for the 2018 data-taking period at {$\sqrt{s} =
  13~\mathrm{TeV}$}'',} CMS Physics Analysis Summary CMS-PAS-LUM-18-002, 2019.

\bibitem{Junk:1999kv}
\hrefCMSnoop {}{T.~Junk, ``{Confidence level computation for combining searches
  with small statistics}'',} \textit{ Nucl. Instrum. Meth. A} \textbf{ 434}
  (1999) 435,
  \href{http://dx.doi.org/10.1016/S0168-9002(99)00498-2}{\doi{10.1016/S0168-9002(99)00498-2}},
  \href{http://www.arXiv.org/abs/hep-ex/9902006}{\texttt{arXiv:hep-ex/9902006}}.

\bibitem{Read_2002}
\hrefCMSnoop {}{A.~L. Read, ``Presentation of search results: the {\CLs}
  technique'',} \textit{ J. Phys. G} \textbf{ 28} (2002) 2693,
  \href{http://dx.doi.org/10.1088/0954-3899/28/10/313}{\doi{10.1088/0954-3899/28/10/313}}.

\bibitem{Cowan:2010js}
\hrefCMSnoop {}{G.~Cowan, K.~Cranmer, E.~Gross, and O.~Vitells, ``{Asymptotic
  formulae for likelihood-based tests of new physics}'',} \textit{ Eur. Phys.
  J. C} \textbf{ 71} (2011) 1554,
  \href{http://dx.doi.org/10.1140/epjc/s10052-011-1554-0}{\doi{10.1140/epjc/s10052-011-1554-0}},
  \href{http://www.arXiv.org/abs/1007.1727}{\texttt{arXiv:1007.1727}}.
  [Erratum: \texttt{doi:10.1140/epjc/s10052-013-2501-z}].

\end{thebibliography}\endgroup
